\documentclass[12pt]{article}

\usepackage{amsfonts}

\begin{document}

\input amssym.def

\input amssym

\centerline{\Large \bf Electronic Fock Spaces: Phase Prefactors and }

\centerline{\Large \bf Hidden Symmetry }

\bigbreak

\centerline {A. I. Panin}

\centerline{ \sl Chemistry Department, St.-Petersburg State University,}

\centerline {University prospect 26, St.-Petersburg 198504, Russia }

\centerline { e-mail: andrej@AP2707.spb.edu }

\bigbreak

{\bf ABSTRACT: }{\small Efficient technique  of manipulation with
phase prefactors  in electronic Fock spaces is developed. Its
power is demonstrated on example of both relatively simple classic
configuration interaction matrix element evaluation  and
essentially more complicated coupled cluster case. Interpretation
of coupled cluster theory in terms of a certain commutative algebra is
given. }

\bigbreak {\bf Key words: }{\small Configuration interaction,
excitation operators, coupled cluster approach, density operators,
commutative algebras}

\bigbreak

{\Large \bf \ Introduction}

\bigbreak

Many years ago in paper \cite {Panin-1} some special
set-theoretical operation ${\Delta}_K$ was introduced. It was used
to reduce manipulations with phase prefactors in the Fock space (
Grassmann algebra) to pure set theoretical ones. More detailed
discussion of this operation was given in \cite {Panin-2},
Appendix A. The main purpose of this paper is to show that the
operation mentioned makes it easy to obtain  very complicated
analytic expressions in effective and uniform manner. In addition,
this technique helps to reveal hidden symmetry of the electronic
Fock space. Fock space equipped with new algebraic structure can
be considered as a natural domain for a certain class of  Post
Hartree-Fock (HF) methods.

In Section I we discuss properties of the operation ${\Delta}_K$
both for the case of molecular spin orbital (MSO) basis and
molecular orbital (MO) basis.

Section II is dedicated to classic configuration interaction (CI)
matrix elements. It does not contain any new results and is included
only as simple but important application of our technique of
manipulations with phase prefactors.

In Section III excitation operators and coupled cluster (CC)
approach  \cite {Bartlett-1} - \cite {Crawford} are discussed. General analytic expression for CI
expansion of CC wave function is given. For the CC CI coefficients
recurrence relations are derived. Analytic expressions for
derivatives of CC CI coefficients with respect to CC amplitudes
are given. CC density operator of the first order necessary for
calculation of molecular properties (see, e.g., \cite {Monkhorst})
is constructed.

Sections IV and V are dedicated to new algebraic structure on the
finite-dimensional electronic Fock spaces.

In Section VI and Appendix  some hints concerning computer
implementation of our approach are given.

\bigbreak \bigbreak

{\Large \bf \ Basic Definitions}

\bigbreak \bigbreak

Let $N=\{1,2,\ldots,n\}$ be the molecular spin-orbital (MSO) index set. On the set
${\cal P}(N)$ of all subsets of $N$ let us consider the operation
(symmetric difference)
$$R\Delta S=(R\cup S)\backslash (R\cap S)\eqno(1)$$
where $R,S\in {\cal P}(N)$. This operation endows ${\cal P}(N)$ with
Abelian group structure with empty set as its unit. Each element of this
group is of order 2 ($R\Delta R=\emptyset $).

Let us define the operation ${\Delta}_K$ as
$${\Delta}_K={\mathop{\Delta}\limits_{k\in K}}\{1,2,\ldots ,k\}\eqno(2)$$
where $K\subset N$. For example, if $K=\{2,4,5,7\}$ then
$${\Delta}_K=\{1,2\}\Delta \{1,2,3,4\}\Delta \{1,2,3,4,5\}\Delta \{1,2,3,4,5,6,7\}=
\{3,4,6,7\}.$$

The mapping
$$\varphi :K\to {\Delta}_K\eqno(3)$$
is a group homomorphism. Indeed, ${\Delta}_{\emptyset}=\emptyset$
and $({\Delta}_K)\Delta ({\Delta}_L)={\Delta}_{K\Delta L}$.

In general case, for  $K=k_1<k_2<\ldots<k_s$
$${\Delta}_K=\cases{
\bigcup\limits_{i=1}^{[{s\over 2}]}\{k_{2i-1}+1,\ldots ,k_{2i}\} &if s is even\cr
\bigcup\limits_{i=0}^{[{s\over 2}]}\{k_{2i}+1,\ldots ,k_{2i+1}\} &if s is odd\cr}
                    \eqno(4)$$

Directly from the definition of operation (1) the following
useful relation may be obtained
$$|K\cap R|+|K\cap S|\equiv |K\cap (R\Delta S)|{\ } (mod{\ } 2)\eqno(5)$$

For a fixed basis set of $n$ orthonormal spin-orbitals the
corresponding finite-dimensional Fock space ${\cal F}_N$ is
spanned by determinants $|R\rangle $ where $R$ runs over all
subsets of the spin-orbital index set $N$. Its $p-$electron sector
${\cal F}_{N,p}$ is spanned by determinants $|R\rangle $ with
$|R|=p$. {\it Basis determinants will be  labelled by subsets
and  all sign conventions connected with their  representation as
the Grassmann product of ordered spin-orbitals will be included in
the definition of the creation-annihilation operators.}

Creation-annihilation operators associated with spin-orbital index $i$ are
defined by the following relations

$$a_i^{\dag}|R\rangle=(1-\zeta_{i,R})(-1)^{\varepsilon}|R\cup \{i\}\rangle\eqno(6a)$$
$$a_i|R\rangle=\zeta_{i,R}(-1)^{\varepsilon}|R\backslash \{i\}\rangle\eqno(6b)$$

where
$$\varepsilon=|\{ 1,2,\ldots,i-1\} \cap R|\eqno(7)$$
is the sign counter and
$$\zeta (I,R)=\cases {1 &if $I\subset R$\cr
                     0 &if $I\not\subset R$\cr }
\eqno(8)$$

is the well-known combinatorial $\zeta$ function of partially
ordered by inclusion set ${\cal P}(N)$. Note that by an abuse of
notation we use symbol $\zeta (i,R)$ instead of $\zeta (\{i\},R)$.
It is pertinent to mention as well that $|R|=|S|$ and $\zeta
(R,S)=1$ imply $R=S$.

The main purpose of this section is to get explicit and easily evaluated
expression for the action of products of the creation-annihilation operators on arbitrary
determinant wave function. We start with calculation of the sum
$|K\cap {\Delta}_ R|+|R\cap {\Delta}_ K|$. From Eq.(5) it readily follows that
for arbitrary $K,R\subset N$
$$|K\cap {\Delta}_ R|+|R\cap {\Delta}_ K|\equiv \sum\limits_{(k,r)\in K\times R}
|\{k\}\cap {\Delta}_ {\{r\}}|+|\{r\}\cap {\Delta}_ {\{k\}}|{\ }
(mod{\ } 2) \eqno(9)$$
But
$$|\{k\}\cap {\Delta}_ {\{r\}}|+|\{r\}\cap {\Delta}_ {\{k\}}| =
\cases {1 &if $k\ne r$\cr
                     2 &if $k=r$\cr }\eqno(10)$$
As a result,
$$|K\cap {\Delta}_ R|+|R\cap {\Delta}_ K|\equiv |K||R|+|K\cap R|{\ } (mod{\ } 2)\eqno(11)$$

Let us return to the creation-annihilation operators and their products.
If $i\not\in R$ then
$$|\{1,2,\ldots,i-1\}\cap R|\equiv |R \cap {\Delta}_i|\equiv
|R|+|\{i\}\cap {\Delta}_{R}|{\ } (mod{\ } 2)\eqno(12a)$$ where
Eq.(11) was used. If $i\in R$ then
$$|\{ 1,2,\ldots,i-1\} \cap R|
\equiv |(R\backslash \{i\}) \cap {\Delta}_i|{\ } (mod{\ } 2)$$ With
the aid of Eq.(11) we can rewrite the last expression as
$$|\{ 1,2,\ldots,i-1\} \cap R|\equiv |R\backslash \{i\}| +
|\{i\}\cap {\Delta}_{R\backslash \{i\}}|$$ $$\equiv |R|+|\{i\}\cap
{\Delta}_{R}| {\ } (mod{\ } 2)\eqno(12b)$$ Now formulas (6a) and
(6b) may be recast as
$$a_i^{\dag}|R\rangle=(-1)^{|R|}(1-\zeta_{i,R})(-1)^{|\{i\}\cap {\Delta}_{R}|}|R\cup \{i\}\rangle\eqno(13a)$$
$$a_i|R\rangle=(-1)^{|R|}\zeta_{i,R}(-1)^{|\{i\}\cap {\Delta}_{R}|}|R\backslash \{i\}\rangle\eqno(13b)$$
Are Eqs.(13a) and (13b) more convenient than Eqs.(6a) and (6b)? In
our opinion in general situation it is more a matter of taste than
convenience. But if the reference determinant $|R \rangle $ is
fixed and it is necessary to calculate the action of different
creation-annihilation operators on this determinant, then it is
possible to find at first the bit vector representing  subset
${\Delta}_R$ and then each time check only one bit of this vector
in contrast to multiple checks if Eq.(7) is used.

 We are interested in expressions  of the type
$a_{i_k}^{\dag}\ldots a_{i_1}^{\dag}a_{j_k}\ldots a_{j_1}|R\rangle
$ involving  {\it particle number preserving} products of the
creation-annihilation operators. Successive application of
Eqs.(13b) and (13a) gives
$$a_{i_k}^{\dag}\ldots a_{i_1}^{\dag}a_{j_k}\ldots a_{j_1}|R\rangle
=$$
$$=\prod\limits_{l=1}^k(1-{\zeta}_{i_l,(R\backslash {J_k}) \cup
I_{l-1}}) \prod\limits_{l=1}^k{\zeta}_{j_l,R\backslash
J_{l-1}}\times (-1)^{\varepsilon}|(R\backslash J_k)\cup I_k
\rangle \eqno(14)$$ where
$$ J_l=\bigcup\limits_{\mu=1}^l\{j_{\mu}\}$$
$$ I_l=\bigcup\limits_{\mu=1}^l\{i_{\mu}\}$$
and
$$\varepsilon = \sum\limits_{l=1}^k \Bigl [ |R\backslash J_{l-1}|+
|(R\backslash J_k)\cup I_{l-1}|+|\{j_l\}\cap {\Delta}_{R\backslash
J_{l-1}}|$$ $$+|\{i_l\}\cap {\Delta}_{(R\backslash J_k)\cup
I_{l-1}}|\Bigr ]\eqno(15)$$

Let us try to simplify the expression for $\varepsilon$. It is
clear  that the right-hand side of Eq.(14) is nonzero only if
$|J_l|=l$ and $|I_l|=l$ for any $l=1,\ldots,k$. As a result,
$$\sum\limits_{l=1}^k \left [ |R\backslash J_{l-1}|+
|(R\backslash J_k)\cup I_{l-1}| \right ] \equiv k {\ } (mod{\ } 2)$$
and the  expression for $\varepsilon$ takes the form
$$\varepsilon = k+
\sum\limits_{l=1}^k \sum\limits_{r=1}^{l-1}|\{j_l\}\cap {\Delta}_{j_r}| +
\sum\limits_{l=1}^k \sum\limits_{r=1}^{k}|\{i_l\}\cap {\Delta}_{j_r}| $$
$$ +\sum\limits_{l=1}^k \sum\limits_{r=1}^{l-1}|\{i_l\}\cap {\Delta}_{i_r}| +
\sum\limits_{l=1}^k |\{j_l\}\cap {\Delta}_R|+ \sum\limits_{l=1}^k
|\{i_l\}\cap {\Delta}_R|\eqno(16)$$ Note that the $R$-dependent
part of this formula is very easy to handle: it is sufficient to
construct the set ${\Delta}_R$ and then just to check if the
indices $j_1,\ldots,j_k$ and $i_1,\ldots,i_k$ belong to this set.
The remainder terms in this formula depend only on the order
relation between indices of the creation-annihilation operators
involved.

Expression (16) takes more simple and compact form if $j_1<\ldots
<j_k$ and $i_1<\ldots <i_k$. Indeed, in this case, as easily
follows from Eq.(4), subsets $\{j_l\}\cap {\Delta}_{J_{l-1}}$ and
$\{i_l\}\cap {\Delta}_{I_{l-1}}$ are empty. As a result,
$$\varepsilon = k+|I_k\cap {\Delta}_{J_k}| +|I_k\cap {\Delta}_R|
+|J_k\cap {\Delta}_R|\eqno(17)$$ If $k=1$ then
$$a_i^{\dag}a_j|R\rangle =(1-{\zeta}_{i,R\backslash \{j\}}){\zeta}_{j,R}
(-1)^{\varepsilon}|(R\backslash \{j\})\cup \{i\}\rangle
\eqno(18)$$ where
$$\varepsilon = 1+|\{i\}\cap {\Delta}_{\{j\}}|+|\{i,j\}\cap {\Delta}_R|
\eqno(19)$$ If $k=2$ then
$$a_i^{\dag}a_j^{\dag}a_la_k|R\rangle =$$
$$=(1-{\zeta}_{i,(R\backslash \{k,l\})
\cup \{j\}})(1-{\zeta}_{j,R\backslash
\{k,l\}}){\zeta}_{l,R\backslash \{k\}}
{\zeta}_{k,R}(-1)^{\varepsilon}|(R\backslash \{k,l\})\cup
\{i,j\}\rangle \eqno(20) $$ where
$$\varepsilon =|\{i\}\cap {\Delta}_{\{j\}}|+|\{i,j,l\}\cap {\Delta}_{\{k\}}|
+|\{i,j\}\cap {\Delta}_{\{l\}}|$$
$$+|\{i,j,k,l\}\cap {\Delta}_R|\eqno(21)$$

In Handy orbital representation \cite {Handy} each MSO index set
is split in its $\alpha$ and $\beta$ components. In particular, in
expression $K\cap {\Delta}_R$ subsets $K,R$ go to
$(K_{\alpha},K_{\beta})$ and $(R_{\alpha},R_{\beta})$,
respectively. Let us introduce the sets $\bar K_{\beta} =
m+K_{\beta}$ and $\bar R_{\beta} = m+R_{\beta}$ where $m$ is the
number of molecular orbitals (MO). We have $K= K_{\alpha}\Delta
\bar K_{\beta}$ and $R= R_{\alpha}\Delta \bar R_{\beta}$ and
$$|K\cap {\Delta}_{R}|\equiv |K_{\alpha}\cap {\Delta}_{R_{\alpha}}|+
|K_{\alpha}\cap {\Delta}_{\bar R_{\beta}}|+
|\bar K_{\beta}\cap {\Delta}_{R_{\alpha}}|+
|\bar K_{\beta}\cap {\Delta}_{\bar R_{\beta}}|\  ({\rm mod}\ 2).$$
It is clear that
$|\bar K_{\beta}\cap {\Delta}_{R_{\alpha}}|=0$ and
$|K_{\alpha}\cap {\Delta}_{\bar R_{\beta}}|\equiv |K_{\alpha}||R_{\beta}|
({\rm mod} \ 2)$. As a result,
$$|K\cap {\Delta}_{R}|\equiv |K_{\alpha}\cap {\Delta}_{R_{\alpha}}|+
|K_{\beta}\cap {\Delta}_{R_{\beta}}|+ |K_{\alpha}||R_{\beta}|({\rm
mod}\ 2). \eqno(22)$$

\bigbreak \bigbreak

\bigbreak

{\Large \bf \ Standard Matrix Elements}

\bigbreak \bigbreak

One-electron operator is of the form
$$h=\sum\limits_{i,j\in N}<i|h|j>a_i^{\dag}a_j\eqno(23)$$
and its action on determinant $|R\rangle $ may be presented as
$$h|R\rangle =-\sum\limits_{j\in R}\sum\limits_{i\in (N\backslash R)\cup \{j\}}
(-1)^{|\{i\}\cap {\Delta}_{\{j\}}|+|\{i,j\}\cap
{\Delta}_R|}<i|h|j>|(R\backslash \{j\})\cup \{i\}\rangle
$$ For matrix element $<S|h|R>$ we have the expression
$$<S|h|R>=-\sum\limits_{j\in R}\sum\limits_{i\in (S\backslash R)\cup \{j\}}
(-1)^{|\{i\}\cap {\Delta}_{\{j\}}|+|\{i,j\}\cap
{\Delta}_R|}<i|h|j>{\zeta}_{(R\backslash \{j\})\cup
\{i\},S}\eqno(24)
$$ Matrix element $<i|h|j>$ is zero if MSO involved are of
different spin. As a result, in MO basis  only cases $j\in
R_{\alpha},i\in S_{\alpha}$ and $j\in R_{\beta},i\in S_{\beta}$
are to be treated. Note that by an abuse of notation we use the
same letters for indices of MSOs and MOs.

\underline {Case '$\alpha \alpha$'}. From Eq.(22) it follows that
$$|(\{k_1\},\emptyset )\cap {\Delta}_R| \equiv |\{k_1\}\cap {\Delta}_{R_{\alpha}}|+
|R_{\beta}|({\rm mod}\ 2)\eqno(25))$$ and
$${\varepsilon}_{\alpha \alpha }=|\{i\}\cap {\Delta}_{\{j\}}|+|\{i,j\}\cap {\Delta}_{R_{\alpha}}|
\eqno(26)$$

\underline {Case '$\beta \beta$'}. We have
$$|(\emptyset ,\{k_1\})\cap {\Delta}_R| \equiv |\{k_1\}\cap {\Delta}_{R_{\beta}}|
({\rm mod}\ 2)\eqno(27)$$ and
$${\varepsilon}_{\beta \beta}=|\{i\}\cap {\Delta}_{\{j\}}|+|\{i,j\}\cap {\Delta}_{R_{\beta}}|
\eqno(28)$$

Thus, in MO basis Eq.(24) becomes
$$<(S_{\alpha},S_{\beta})|h|(R_{\alpha},R_{\beta})>=$$
$$-{\zeta}_{R_{\beta},S_{\beta}}\sum\limits_{j\in R_{\alpha}}
\sum\limits_{i\in (S_{\alpha}\backslash R_{\alpha})\cup \{j\}}
(-1)^{{\varepsilon}_{\alpha \alpha }}<i|h|j>
{\zeta}_{({R_{\alpha}}\backslash \{j\})\cup \{i\},S_{\alpha}} $$
$$-{\zeta}_{R_{\alpha},S_{\alpha}}\sum\limits_{j\in R_{\beta}}
\sum\limits_{i\in (S_{\beta}\backslash R_{\beta})\cup \{j\}}
(-1)^{{\varepsilon}_{\beta \beta}}<i|h|j>
{\zeta}_{({R_{\beta}}\backslash \{j\})\cup
\{i\},S_{\beta}}\eqno(29)
$$

Two-electron operator is of the form
$$g=\sum\limits_{i,j,k,l\in N}<ij|kl>a_i^{\dag}a_j^{\dag}a_la_k\eqno(30)$$
and its matrix element $<S|g|R>$ may be written as
$$<S|g|R>=
\sum\limits_{k,l\in R\atop{(k\ne l})}\sum\limits_{i,j\in
(S\backslash R)\cup \{k,l\}\atop{(i\ne
j)}}(-1)^{\varepsilon}<ij|kl> {\zeta}_{(R\backslash \{k,l\})\cup
\{i,j\},S}\eqno(31)$$ where ${\varepsilon}$ is given by Eq.(21).

Two-electron integrals $<ij|kl>$ equal to zero if MSOs with indices
$i,k$ or with indices $j,l$ are of different spins. In MO basis two-electron
integrals are written as $(ik|jl)$ and there are four cases to be analyzed.

\underline {Case '$\alpha \alpha \alpha \alpha$'}. All orbitals are of the
same spin $\alpha $ and in this case
$$|(\{k_1,k_2\},\emptyset )\cap {\Delta}_R| \equiv |\{k_1,k_2\}\cap {\Delta}_{R_{\alpha}}|+
2|R_{\beta}|({\rm mod}\ 2)\eqno(32)$$ and
$${\varepsilon}_{\alpha \alpha \alpha \alpha}=
|\{l\}\cap {\Delta}_{\{k\}}|+|\{j\}\cap {\Delta}_{\{k\}}|
+|\{i\}\cap {\Delta}_{\{k\}}|+|\{j\}\cap {\Delta}_{\{l\}}|$$ $$+
|\{i\}\cap {\Delta}_{\{l\}}|+|\{i\}\cap {\Delta}_{\{j\}}|
+|\{i,j\}\cap {\Delta}_{R_{\alpha}}|+ |\{k,l\}\cap
{\Delta}_{R_{\alpha}}|\eqno(33)$$

\underline {Case '$\alpha \alpha \beta \beta$'}. Orbitals with indices
$i,k$ are of $\alpha $ spin and with indices $j,l$ are of $\beta $ spin. We have
$$|(\{k_1\},\{k_2\})\cap {\Delta}_R| \equiv |\{k_1\}\cap {\Delta}_{R_{\alpha}}|+
|\{k_2\}\cap {\Delta}_{R_{\beta}}|+|R_{\beta}|({\rm mod}\
2)\eqno(34)$$ and
$${\varepsilon}_{\alpha \alpha \beta \beta}=
|\{i\}\cap {\Delta}_{\{k\}}|+|\{j\}\cap {\Delta}_{\{l\}}|+
|\{i,k\}\cap {\Delta}_{R_{\alpha}}|+ |\{j,l\}\cap
{\Delta}_{R_{\beta}}|\eqno(35)$$

\underline {Case '$\beta \beta \alpha \alpha $'}. Orbitals with indices
$i,k$ are of $\beta $ spin and with indices $j,l$ are of $\alpha $ spin. We have
$${\varepsilon}_{\beta \beta \alpha \alpha }=
|\{i\}\cap {\Delta}_{\{k\}}|+|\{j\}\cap {\Delta}_{\{l\}}|+
|\{i,k\}\cap {\Delta}_{R_{\beta}}|+ |\{j,l\}\cap
{\Delta}_{R_{\alpha}}|\eqno(36)$$

\underline {Case '$\beta \beta \beta \beta$'}. All orbitals are of the
same spin $\beta $ and in this case
$$|(\emptyset ,\{k_1,k_2\})\cap {\Delta}_R| \equiv |\{k_1,k_2\}\cap {\Delta}_{R_{\beta}}|
({\rm mod}\ 2)\eqno(37)$$ and
$${\varepsilon}_{\beta \beta \beta \beta}=
|\{l\}\cap {\Delta}_{\{k\}}|+|\{j\}\cap {\Delta}_{\{k\}}|
+|\{i\}\cap {\Delta}_{\{k\}}|+|\{j\}\cap {\Delta}_{\{l\}}|$$ $$+
|\{i\}\cap {\Delta}_{\{l\}}|+|\{i\}\cap {\Delta}_{\{j\}}|
+|\{i,j\}\cap {\Delta}_{R_{\beta}}|+ |\{k,l\}\cap
{\Delta}_{R_{\beta}}|\eqno(38)$$

In MO basis the expression (31) takes the form

$$<(S_{\alpha},S_{\beta})|g|(R_{\alpha},R_{\beta})>=$$
$$
{\zeta}_{R_{\beta},S_{\beta}} \sum\limits_{k,l\in
R_{\alpha}\atop{(k\ne l)}}\sum\limits_{i,j\in
(S_{\alpha}\backslash R_{\alpha})\cup \{k,l\}\atop{(i\ne j)}}
(-1)^{{\varepsilon}_{\alpha \alpha \alpha \alpha }}(ik|jl)
{\zeta}_{(R_{\alpha}\backslash \{k,l\})\cup \{i,j\},S_{\alpha}}$$
$$+\sum\limits_{k\in R_{\alpha}\atop {l\in R_{\beta}}}
\sum\limits_{i\in (S_{\alpha}\backslash R_{\alpha})\cup \{k\}
\atop {j\in (S_{\beta}\backslash R_{\beta})\cup \{l\}}}
(-1)^{{\varepsilon}_{\alpha \alpha \beta \beta }}(ik|jl)
{\zeta}_{(R_{\alpha}\backslash \{k\})\cup \{i\},S_{\alpha}}
{\zeta}_{(R_{\beta}\backslash \{l\})\cup \{j\},S_{\beta}}$$
$$+\sum\limits_{k\in R_{\beta}\atop {l\in R_{\alpha}}}
\sum\limits_{i\in (S_{\beta}\backslash R_{\beta})\cup \{k\}
\atop {j\in (S_{\alpha}\backslash R_{\alpha})\cup \{l\}}}
(-1)^{{\varepsilon}_{\beta \beta \alpha \alpha }}(ik|jl)
{\zeta}_{(R_{\alpha}\backslash \{l\})\cup \{j\},S_{\alpha}}
{\zeta}_{(R_{\beta}\backslash \{k\})\cup \{i\},S_{\beta}}$$
$$+
{\zeta}_{R_{\alpha},S_{\alpha}} \sum\limits_{k,l\in
R_{\beta}\atop{(k\ne l)}}\sum\limits_{i,j\in (S_{\beta}\backslash
R_{\beta})\cup \{k,l\}\atop{(i\ne j)}} (-1)^{{\varepsilon}_{\beta
\beta \beta \beta}}(ik|jl) {\zeta}_{(R_{\beta}\backslash
\{k,l\})\cup \{i,j\},S_{\beta}}\eqno(39)$$ To reduce the dimension
of configuration interaction (CI) space one may turn to the notion
of the active space. Let us suppose that each determinant under
consideration is presented in the form
$$(R_{\alpha},R_{\beta})=(I\cup R_{\alpha}^a,I\cup R_{\beta}^a)\eqno(40)$$
where $I=\{1,2,\ldots,n_i\}$ is the set of the so-called inactive MO indices
that occur in each determinant with the occupancy 2.

From Eq.(29) it easily follows that either $i,j\in I$ or
$i,j\not\in I$. As a result,
$$<(I\cup S_{\alpha}^a,I\cup S_{\beta}^a)|h|(I\cup R_{\alpha}^a,I\cup R_{\beta}^a)>=$$
$${\zeta}_{R_{\alpha}^a,S_{\alpha}^a}{\zeta}_{R_{\beta}^a,S_{\beta}^a}
\sum\limits_{i\in I}2<i|h|i>$$
$$-{\zeta}_{R_{\beta}^a,S_{\beta}^a}\sum\limits_{j\in R_{\alpha}^a}
\sum\limits_{i\in (S_{\alpha}^a\backslash R_{\alpha}^a)\cup \{j\}}
(-1)^{{\varepsilon}_{\alpha \alpha }}<i|h|j>
{\zeta}_{({R_{\alpha}^a}\backslash \{j\})\cup \{i\},S_{\alpha}^a} $$
$$-{\zeta}_{R_{\alpha}^a,S_{\alpha}^a}\sum\limits_{j\in R_{\beta}^a}
\sum\limits_{i\in (S_{\beta}^a\backslash R_{\beta}^a)\cup \{j\}}
(-1)^{{\varepsilon}_{\beta \beta}}<i|h|j>
{\zeta}_{({R_{\beta}^a}\backslash \{j\})\cup
\{i\},S_{\beta}^a}\eqno(41) $$ In the same manner Eq.(39) may be
treated. We have
$$<(I\cup S_{\alpha}^a,I\cup S_{\beta}^a)|g|(I\cup R_{\alpha}^a,I\cup R_{\beta}^a)>=$$
$${\zeta}_{R_{\alpha}^a,S_{\alpha}^a}{\zeta}_{R_{\beta}^a,S_{\beta}^a}
\sum\limits_{i,j\in I}2\left [2(ii|jj)-(ij|ji)\right ]$$
$$+
{\zeta}_{R_{\beta}^a,S_{\beta}^a} \sum\limits_{k,l\in
R_{\alpha}^a\atop{(k\ne l)}}\sum\limits_{i,j\in
(S_{\alpha}^a\backslash R_{\alpha}^a)\cup \{k,l\}\atop{(i\ne j)}}
(-1)^{{\varepsilon}_{\alpha \alpha \alpha \alpha }}(ik|jl)
{\zeta}_{(R_{\alpha}^a\backslash \{k,l\})\cup
\{i,j\},S_{\alpha}^a}$$
$$+\sum\limits_{k\in R_{\alpha}^a\atop {l\in R_{\beta}^a}}
\sum\limits_{i\in (S_{\alpha}^a\backslash R_{\alpha}^a)\cup \{k\}
\atop {j\in (S_{\beta}^a\backslash R_{\beta}^a)\cup \{l\}}}
(-1)^{{\varepsilon}_{\alpha \alpha \beta \beta }}(ik|jl)
{\zeta}_{(R_{\alpha}^a\backslash \{k\})\cup \{i\},S_{\alpha}^a}
{\zeta}_{(R_{\beta}^a\backslash \{l\})\cup \{j\},S_{\beta}^a}$$
$$+\sum\limits_{k\in R_{\beta}^a\atop {l\in R_{\alpha}^a}}
\sum\limits_{i\in (S_{\beta}^a\backslash R_{\beta}^a)\cup \{k\}
\atop {j\in (S_{\alpha}^a\backslash R_{\alpha}^a)\cup \{l\}}}
(-1)^{{\varepsilon}_{\beta \beta \alpha \alpha }}(ik|jl)
{\zeta}_{(R_{\alpha}^a\backslash \{l\})\cup \{j\},S_{\alpha}^a}
{\zeta}_{(R_{\beta}^a\backslash \{k\})\cup \{i\},S_{\beta}^a}$$
$$+
{\zeta}_{R_{\alpha}^a,S_{\alpha}^a} \sum\limits_{k,l\in
R_{\beta}^a\atop{(k\ne l)}}\sum\limits_{i,j\in
(S_{\beta}^a\backslash R_{\beta}^a)\cup \{k,l\}\atop{(i\ne j)}}
(-1)^{{\varepsilon}_{\beta \beta \beta \beta}}(ik|jl)
{\zeta}_{(R_{\beta}^a\backslash \{k,l\})\cup
\{i,j\},S_{\beta}^a}\eqno(42)$$

\bigbreak \bigbreak

{\Large \bf \  Excitation Operators}

\bigbreak \bigbreak

The operator performing $k$-fold excitation  from some fixed single determinant
reference state $|R>$ is defined as
$$\mathfrak{a}_{J_k}^{I_k}=
a_{i_k}^{\dag}\ldots a_{i_1}^{\dag}a_{j_k}\ldots
a_{j_1}\eqno(43)$$ where $J_k=\{j_1<\ldots<j_k\}\subset R$ and
$I_k=\{i_1<\ldots<i_k\}\subset N\backslash R$. We have
$$\mathfrak{a}_{J_k}^{I_k}|R>=(-1)^{k+|I_k\cap {\Delta}_{J_k}|+|I_k\cap {\Delta}_R|+
|J_k\cap {\Delta}_R|}|(R\backslash J_k)\cup I_k>$$
Product
$$\mathfrak{a}_{J_{k_{\mu}}\ldots J_{k_1}}^{I_{k_{\mu}}\ldots I_{k_1}}=
\mathfrak{a}_{J_{k_{\mu}}}^{I_{k_{\mu}}}\ldots
\mathfrak{a}_{J_{k_1}}^{I_{k_1}}\eqno(44)$$ gives non-vanishing
result when applied to $|R>$ if and only if families
$I_{k_1},\ldots,I_{k_{\mu}}$ and $J_{k_1},\ldots,J_{k_{\mu}}$
include {\it mutually disjoint} subsets. After simple manipulations we
get
$$\mathfrak{a}_{J_{k_{\mu}}\ldots J_{k_1}}^{I_{k_{\mu}}\ldots I_{k_1}}|R>=
(-1)^{\varepsilon}|(R\backslash J)\cup I>\eqno(45)$$ where
$${\varepsilon}=\sum\limits_{i=1}^{\mu}k_i+\sum\limits_{i=1}^{\mu}|I_{k_i}\cap {\Delta}_{J_{k_i}}|
+|I\cap {\Delta}_R|+|J\cap {\Delta}_R|
+\sum\limits_{i>j}|(I_{k_i}\cup J_{k_i})\cap
{\Delta}_{(I_{k_j}\cup J_{k_j})}| \eqno(46)$$ and
$I=\bigcup\limits_{i=1}^{\mu}I_{k_i},J=\bigcup\limits_{i=1}^{\mu}J_{k_i}$.

The expression for $\varepsilon $ can be simplified if we suppose
that the reference determinant $|R\rangle $ is built on the first
$p\ $ spin-orbitals that is $|R\rangle =|\{1,2,\ldots,p\}\rangle $. We,
however, do not presuppose any special choice of indices of
occupied HF MSOs in our further discussion.

{\bf Definition 1.} Family $\{J_i\}_{i=1}^{\mu}$ is a set
 partition of $J$ into $\mu $ blocks if

(1) $J_i\ne\emptyset $ for all $i=1,2,\ldots,\mu$;

(2) $J_i\cap J_j=\emptyset $ for all $i\ne j$;

(3) $J=\bigcup\limits_{i=1}^{\mu} J_i$.

Normally set partition has its blocks listed in increasing order
of smallest element in each block. We will call such partitions
block  ordered ones, reserving the term 'set partitions' for
arbitrary families with properties (1) - (3) of Definition 1.

If $|J|=k$ then the number of all possible block ordered set
partitions of $J$ is equal to the $k$th Bell number $B_k$ (see,
e.g., \cite {Stanley}).

{\bf Definition 2.} Family $\{(J_i,I_i)\}_{i=1}^{\mu})$ is a
consistent set partition of a pair of subsets $(J,I)$ if

(1) $\{J_i\}_{i=1}^{\mu}$ is  a block ordered set partition of
$J$;

(2) $\{I_i\}_{i=1}^{\mu}$ is a set partition of $I$;

(3) $|J_i|=|I_i|$ for all $i=1,2,\ldots,\mu$.

Excitation operators appear explicitly in coupled cluster theory
\cite {Bartlett-1} - \cite {Crawford} where the exponential anzatz of wave function is used:
$$|\Psi> = e^X|R>.\eqno(47)$$
Here
$$X=\sum\limits_{k=1}^l\sum\limits_{J_k\subset R \atop {I_k\subset N\backslash R}}^{(k)}
t_{J_k}^{I_k}\mathfrak{a}_{J_k}^{I_k},\eqno(48)$$

is the excitation operator associated with chosen reference
determinant $|R>$, and $l\le p$ is the maximal excitation order.

Let us consider the standard expansion
$$e^X=\sum\limits_{\mu\ge 0}\frac{1}{\mu !}X^{\mu}\eqno(49)$$
and try to present the expression for the restriction of $X^{\mu}$
on one-dimensional reference subspace $\mathbb C |R>$   in a form
convenient for further analysis. We have
$$X^{\mu}|R>=\sum\limits_{k=\mu}^{min(p,\mu l)} \sum\limits_{J\subset R \atop {I\subset N\backslash R}}^{(k)}
\sum\limits_{J_{k_1},\ldots,J_{k_{\mu}}\subset J \atop
{I_{k_1},\ldots,I_{k_{\mu}}\subset I}}^{(k_1+\ldots k_{\mu}=k)}
t_{J_{k_{\mu}}}^{I_{k_{\mu}}}\ldots t_{J_{k_1}}^{I_{k_1}}
\mathfrak{a}_{J_{k_{\mu}}\ldots J_{k_1}}^{I_{k_{\mu}}\ldots
I_{k_1}}|R>\eqno(50)$$ where the inner sum goes over families that
determine consistent set partitions of subsets $J$ and $I$ with
$1\le k_i\le l$. Summation of both sides of the last equality over
$\mu$ gives
$$e^X|R>=|R>+\sum\limits_{\mu=1}^p
\sum\limits_{k=\mu}^{min(p,\mu l)}\sum\limits_{J\subset R \atop
{I\subset N\backslash R}}^{(k)} \frac{1}{\mu !}
\sum\limits_{J_{k_1},\ldots,J_{k_{\mu}}\subset J \atop
{I_{k_1},\ldots,I_{k_{\mu}}\subset I}}^{(k_1+\ldots k_{\mu}=k)}
t_{J_{k_{\mu}}}^{I_{k_{\mu}}}\ldots t_{J_{k_1}}^{I_{k_1}}
\mathfrak{a}_{J_{k_{\mu}}\ldots J_{k_1}}^{I_{k_{\mu}}\ldots
I_{k_1}}|R>\eqno(51)$$

The right-hand side of Eq.(51) contains  similar terms. Indeed, if
$\mu$ is fixed and some family $(\{J_{k_i}\}_{i=1}^{\mu},
\{I_{k_i}\}_{i=1}^{\mu})$ is selected then any permutation
$\sigma\in S_{\mu}$ generates new family
$(\{J_{k_{\sigma(i)}}\}_{i=1}^{\mu},\{I_{k_{\sigma(i)}}\}_{i=1}^{\mu})$.
Of $\mu !$ families thus obtained let us choose one and put
$${\varepsilon}_{\sigma}=
\sum\limits_{i>j}|(I_{k_{\sigma(i)}}\cup J_{k_{\sigma(i)}}) \cap
{\Delta}_{(I_{k_{\sigma(j)}}\cup J_{k_{\sigma(j)}})}| \eqno(52)$$
Then the summation over $\mu !$ families on the right-hand side of
Eq.(51) is reduced to calculation of the following sum
$$s_{\mu}(\{J_{k_i}\}_{i=1}^{\mu},\{I_{k_i}\}_{i=1}^{\mu})=
\frac{(-1)^{{\varepsilon}_0}}{\mu !}\sum\limits_{\sigma \in
S_{\mu}} (-1)^{{\varepsilon}_{\sigma}}\eqno(53)$$ where
$${\varepsilon}_0=\sum\limits_{i=1}^{\mu}k_i+\sum\limits_{i=1}^{\mu}|I_k\cap {\Delta}_{J_k}|
+|I\cap {\Delta}_R|+|J\cap {\Delta}_R|\eqno(54)$$ But from Eq.(11)
it easily follows that
$$|(I_{k_i}\cup J_{k_i})\cap {\Delta}_{(I_{k_j}\cup J_{k_j})}|+
|(I_{k_j}\cup J_{k_j})\cap {\Delta}_{(I_{k_i}\cup J_{k_i})}|\equiv
\ 0 (mod\ 2).$$ As a result,
$s_{\mu}(\{J_{k_i}\}_{i=1}^{\mu},\{I_{k_i}\}_{i=1}^{\mu})=(-1)^{\varepsilon}$
and final expression for $e^X|R>$ is
$$e^X|R>=\sum\limits_{k=0}^p\sum\limits_{J\subset R \atop {I\subset N\backslash R}}^{(k)}
 T_k^l(J,I)|(R\backslash J)\cup I>,\eqno(55)$$ where
$$T_k^l(J,I)=\sum\limits_{\mu\in M_k}\sum\limits_{J_{k_1},\ldots,J_{k_{\mu}}\subset
J \atop {I_{k_1},\ldots,I_{k_{\mu}}\subset I}}^{(k_1+\ldots
k_{\mu}=k)} (-1)^{\varepsilon}t_{J_{k_1}}^{I_{k_1}}\ldots
t_{J_{k_{\mu}}}^{I_{k_{\mu}}}\eqno(56a)$$ and
$$T_0^l(\emptyset ,\emptyset )=1\eqno(56b)$$
 In Eq.(56) the index set $M_k$ is defined as
 $$M_k=\{\mu:\left [\mu,min(p,\mu l)\right ]\ni k\}$$
 It is easy to see that $k$ is the maximal element of $M_k$. Its minimal element
 may be found from the inequality
 $$k=l\cdot \left [\frac{k}{l}\right ] +r \le \mu\cdot l$$
 where $0\le r<l$, and is given by the following relation
$${\mu}_{min}=\cases{[\frac {k}{l}] &if $k\equiv 0 (mod \ l)$\cr
           [\frac {k}{l}]+1 &if $k\not\equiv 0 (mod \ l)$\cr}\eqno(57)$$

This ${\mu}_{min}$ corresponds to the minimal possible number of
blocks of sizes not greater than $l$ (see Appendix). {\it Thus, the
summation in Eq.(56a) goes actually over all consistent set
partitions of the pair $(J,I)$ such that each block size does not
exceed $l$, and $\varepsilon$ is defined by Eq.(46).}

At this stage it seems reasonable to introduce new CC amplitudes
that differ from the initial ones defined by Eq.(48), only by
phase prefactors:
$$t_{J}^{I} \to (-1)^{|J|+|I\cap {\Delta}_J|+|I\cap {\Delta}_R|+|J\cap
{\Delta}_R|}t_J^I\eqno(58)$$ By an abuse of notation for phase modified CC
amplitudes we will use the same symbols as for the initial ones.
With new definition of CC amplitudes Eqs.(55) and (56) stay
unchanged except for ${\varepsilon}$ in Eq.(56) that becomes
$${\varepsilon}=
\sum\limits_{i>j}|(I_{k_i}\cup J_{k_i})\cap {\Delta}_{(I_{k_j}\cup
J_{k_j})}| \eqno(59)$$

The expression (55) establishes a connection between CC and CI
expansions in most general case. It is pertinent to mention that
in CC expansion all determinants are involved, even if the CC
operator (48) is of low excitation order.

Now we are ready  to  get the expression for the first order
density operator
$${\rho }_R^{CC}= \frac{1}{p!}c^{p-1}\left
[e^X|R><R|e^X\right ],\eqno(60)$$ where the contraction operator
is defined by
$$\frac{1}{p!}c^{p-1}|R\rangle \langle S|=\frac {1}{p}
\cases{ \sum\limits_{j \in R}|j \rangle \langle j| &if S=R\cr
-(-1)^{\delta}|j\rangle \langle i| &if $S=(R\backslash \{j\})\cup
\{i\}$\cr 0 &if $|S\cap R|<p-1$ \cr } \eqno(61)$$ with
$$\delta= |\{i\}\cap {\Delta}_{\{j\}}|+|\{i,j\}\cap {\Delta
}_R|$$ and $i\ne j$.

After simple manipulations we come to the following expression:
$$p{\rho }_R^{CC}=
\sum\limits_{k=0}^p\sum\limits_{J\subset R\atop {I\subset
N\backslash R}}^{(k)}\left [ T_k^l(J,I)\right ]^2
\sum\limits_{j\in R\backslash J \cup I}|j\rangle \langle j|$$
$$-\sum\limits_{k=1}^p\sum\limits_{J\subset R\atop{I\subset N\backslash
R}}^{(k)}T_k^l(J,I)\sum\limits_{j\in R\backslash J\atop {i\in J}}
(-1)^{{\gamma}_1+{\gamma}_2}T_k^l((J\backslash \{i\})\cup
\{j\},I)|j\rangle \langle i|$$
$$-\sum\limits_{k=0}^{p-1}\sum\limits_{J\subset R\atop{I\subset N\backslash
R}}^{(k)}T_k^l(J,I)\sum\limits_{j\in R\backslash J\atop {i\in
N\backslash R\backslash I}}
(-1)^{{\gamma}_1+{\gamma}_2}T_{k+1}^l(J\cup \{j\},I\cup
\{i\})|j\rangle \langle i|$$
$$-\sum\limits_{k=1}^p\sum\limits_{J\subset R\atop{I\subset N\backslash
R}}^{(k)}T_k^l(J,I)\sum\limits_{j\in I\atop {i\in J}}
(-1)^{{\gamma}_1+{\gamma}_2}T_{k-1}^l(J\backslash
\{i\},I\backslash \{j\})|j\rangle \langle i|$$
$$-\sum\limits_{k=1}^p\sum\limits_{J\subset R\atop{I\subset N\backslash
R}}^{(k)}T_k^l(J,I)\sum\limits_{j\in I\atop {i\in N\backslash
R\backslash I}} (-1)^{{\gamma}_1+{\gamma}_2}T_k^l(J,(I\backslash
\{j\})\cup \{i\})|j\rangle \langle i|\eqno(62a)$$ where
$${\gamma}_1= |\{i\}\cap
{\Delta}_{\{j\}}|, \ {\gamma}_2= |\{i,j\}\cap {\Delta
}_{(R\backslash J)\cup I}|\eqno(62b)$$ or, after collecting
similar terms,
$$p{\rho }_R^{CC}=
\sum\limits_{j\in R}|j\rangle \langle j|
\sum\limits_{k=0}^p\sum\limits_{J\subset R\backslash \{j\}\atop
{I\subset N\backslash R}}^{(k)}\left [ T_k^l(J,I)\right ]^2$$
$$+\sum\limits_{j\in N\backslash R}|j\rangle \langle
j|\sum\limits_{k=1}^p\sum\limits_{J\subset
R}^{(k)}\sum\limits_{I\subset N\backslash R\backslash
\{j\}}^{(k-1)}\left [ T_k^l(J,I\cup\{j\})\right ]^2$$
$$-\sum\limits_{i,j\in R\atop{(i\ne j)}}|j\rangle \langle
i|\left [\sum\limits_{k=1}^p\sum\limits_{J\subset R\backslash
\{i,j\}}^{(k-1)}\sum\limits_{I\subset N\backslash
R}^{(k)}(-1)^{{\gamma}_1+{\gamma}_2
}T_k^l(J\cup\{i\},I)T_k^l(J\cup \{j\},I)\right ]$$
$$-\sum\limits_{j\in R\atop{i\in N\backslash R}}|j\rangle \langle
i|\left [ \sum\limits_{k=0}^{p-1}\sum\limits_{J\subset R\backslash
\{j\}\atop{I\subset N\backslash R\backslash
\{i\}}}^{(k)}(-1)^{{\gamma}_2}T_k^l(J,I)T_{k+1}^l(J\cup
\{j\},I\cup \{i\})\right ]$$
$$+\sum\limits_{j\in N\backslash R\atop{i\in R}}|j\rangle \langle
i|\left [ \sum\limits_{k=1}^p\sum\limits_{J\subset R\backslash
\{i\}\atop{I\subset N\backslash R\backslash
\{j\}}}^{(k-1)}(-1)^{{\gamma}_2}T_k^l(J\cup \{i\},I\cup \{j\}
)T_{k-1}^l(J,I)\right ]$$
$$+\sum\limits_{i,j\in N\backslash R\atop{(i\ne j)}}|j\rangle \langle
i|\left [\sum\limits_{k=1}^p\sum\limits_{J\subset
R}^{(k)}\sum\limits_{I\subset N\backslash R \backslash
\{i,j\}}^{(k-1)} (-1)^{{\gamma}_1+{\gamma }_2}T_k^l(J,I\cup
\{j\})T_k^l(J,I\cup \{i\})\right ]\eqno(63)$$

In the analogous manner the expression for the second order CC
density operator may be obtained.

For further analysis it is convenient to introduce the
coefficients
$$\left [T_{k,q}^l(\tau)\right ]_{JI}=\sum\limits_{\mu={\mu}_{min}}^k\frac{\mu !}{(\mu+q)!}
\sum\limits_{J_{k_1},\ldots,J_{k_{\mu}}\subset J \atop
{I_{k_1},\ldots,I_{k_{\mu}}\subset I}}^{(k_1+\ldots +k_{\mu}=k)}
(-1)^{\varepsilon}t_{J_{k_1}}^{I_{k_1}}\ldots
t_{J_{k_{\mu}}}^{I_{k_{\mu}}}\eqno(64)$$ where ${\mu}_{min}$ is
defined by Eq.(58) and $\tau$ is the vector (matrix) of CC
amplitudes. It is clear that $\left [T_{k,0}^l(\tau)\right
]_{JI}=T_k^l(J,I)$, and that $\left [T_{0,q}^l(\tau)\right
]_{JI}=\frac{1}{q!}$.

Let us present the coefficient $\left [T_{k,q}^l(\tau)\right
]_{JI}$ in the form
$$\left [T_{k,q}^l(\tau)\right ]_{JI}=\int\limits_0^1\frac{d}{d\lambda}
\left [T_{k,q}^l(\lambda \tau)\right ]_{JI}d\lambda=$$ $$=
\sum\limits_{k_*=1}^{min(k,l)}\sum\limits_{J_*\subset
R\atop{I_*\subset N\backslash
R}}^{(k_*)}\int\limits_0^1\frac{\partial}{\partial (\lambda
t_{J_*}^{I_*})}\left [T_{k,q}^l(\lambda\tau)\right
]_{JI}t_{J_*}^{I_*}d\lambda\eqno(65)$$

Calculation of the derivatives of $\left [T_{k,q}^l(\tau)\right
]_{JI}$ with respect to (phase modified) CC amplitudes we start
with the analysis of phase prefactor. To this end let us recast
Eq.(59), using Eq.(11), as
$$\varepsilon =\sum\limits_{i>j\atop{(i,j\ne i_*)}}|(I_{k_i}\cup
J_{k_i})\cap {\Delta}_{(I_{k_j}\cup J_{k_j})}|$$
$$+\sum\limits_{i}^{\mu}|(I_{k_i}\cup
J_{k_i})\cap {\Delta}_{(I_{k_{i_*}}\cup J_{k_{i_*}})}|+
|(I_{k_{i_*}}\cup J_{k_{i_*}})\cap {\Delta}_{(I_{k_{i_*}}\cup
J_{k_{i_*}})}|.$$ Taking into account that
$$|(I_{k_{i_*}}\cup J_{k_{i_*}})\cap {\Delta}_{(I_{k_{i_*}}\cup
J_{k_{i_*}})}|\equiv \left [ \frac{2k_{i_*}+1}{2}\right ]\equiv
k_{i_*} (mod\ 2)$$ we can write the following elegant expression
$$\int\limits_0^1\frac{\partial}{\partial (\lambda
t_{J_*}^{I_*})}\left [T_{k,q}^l(\lambda\tau)\right
]_{JI}d\lambda=$$
$$=\cases { (-1)^{{\eta}(J_*,I_*)}\left [
T_{k-k_*,q+1}^l(\tau)\right ] _{J\backslash J_*I\backslash I_*}
&if $J \supset J_*$ and $I\supset I_*$ \cr 0 &if $J \not\supset
J_*$ or $I\not\supset I_*$\cr }\eqno(66)$$ where
$${\eta}(J_*,I_*)=k_*+|(J_*\cup
I_*)\cap{\Delta}_{(J\cup I)}|.\eqno(67)$$

Now Eq.(65) may be recast in the form of the recurrence relation:
$$\left [T_{k,q}^l(\tau)\right ]_{JI}=\sum\limits_{k_*=1}^{min(k,l)}\sum\limits_{J_*\subset
J\atop{I_*\subset I}}^{(k_*)}(-1)^{{\eta}(J_*,I_*)}\left [
T_{k-k_*,q+1}^l(\tau)\right ] _{J\backslash J_*I\backslash I_*}
t_{J_*}^{I_*}\eqno(68)$$
CC CI coefficients correspond to the case $q=0$ in this equation. The
recursion can go either to the vacuum $(k-k_*=0)$ or to small values of
$k-k_*$ for which the required CC CI coefficients can be evaluated directly.

\bigbreak \bigbreak
\newpage

{\Large \bf \ Structure of Commutative Algebra  on}

{\Large \bf \ p - Electron Sector of the Fock Space }

\bigbreak \bigbreak

Let us denote by the symbol ${\cal A}_R^p$ the $p-$electron sector
${\cal F}_{N,p}$ of
the Fock space where some determinant $|R\rangle$ is selected. In more formal
terms ${\cal A}_R^p$ is {\it a pointed space} that is the pair
$({\cal F}_{N,p}\ ,|R\rangle)$. The selected point $|R\rangle$ is referred to
as  either the HF vacuum or the HF reference state.

${\cal A}_R^p$ is spanned by the basis vectors
$$e_J^I(R)=|(R\backslash J)\cup I\rangle\eqno(69)$$
labelled by subsets $J\subset
R,I\subset N\backslash R$ with $|J|=|I|$.
{\it Note that} $e_J^I(R)$ {\it is nothing more than a special convenient
notation for the basis determinant in} ${\cal F}_{N,p}$.

If $S=(R\backslash J)\cup I$ then it is easy to ascertain that
$$e_{J'}^{I'}(S)=e_{(J\backslash I_1')\cup {J_1'}}^
{(I\backslash {J_2')}\cup {I_2'}}(R)\eqno(70)$$
where
$$J_1'=J'\cap (R\backslash J),\  J_2'=J'\cap I\eqno(71a)$$
and
$$I_1'=I'\cap J,\  I_2'=I'\cap (N\backslash R\backslash I).\eqno(71b)$$
In particular, $e_{\emptyset}^{\emptyset}(S)=e_J^I(R)$ and $e_I^J(S)=
e_{\emptyset}^{\emptyset}(R)$.

Therefore, arbitrary finite linear combinations of vectors corresponding to
different vacuum states are allowed since with the aid of Eqs.(70) - (71) such
combinations are easily reduced to some common vacuum state.

We furthermore assume that the vacuum vector $|R\rangle$ is fixed
and, by an abuse of notation, for basis vectors (69) we will use
the symbol $e_J^I$.

${\cal A}_R^p$ is  Hermitean space
with scalar product
$$\langle x | y\rangle =\sum\limits_{k=0}^p\sum\limits_{J\subset R \atop {I\subset N\backslash R}}^{(k)}
(x_J^I)^*y_J^I\eqno(72a)$$ and norm
$$\|x\|=\left [\sum\limits_{k=0}^p\sum\limits_{J\subset R \atop {I\subset N\backslash R}}^{(k)}
|x_J^I|^2\right ]^{\frac{1}{2}}.\eqno(72b)$$
Subspaces ${\cal A}_R^l$ of the dimension
$$dim\  {\cal
A}_R^l=\sum\limits_{k=0}^l {n\choose k}{{n-p}\choose k},\eqno(73)$$ spanned
by the basis vectors $e_J^I$ with  $1\le |J|=|I|\le l$, form the
following chain
$${\cal A}_R^0\subset {\cal A}_R^1\subset \ldots \subset {\cal A}_R^p\eqno(74a)$$
where
$${\cal A}_R^0={\mathbb C}e_{\emptyset}^{\emptyset}.\eqno(74b)$$
For each pair $(k,l)$ with $k\le l$ we introduce subspace ${\cal
W}_R^{(k,l)}$ spanned by vectors $e_J^I$ with $k\le |J|=|I|\le l$.
It is clear that ${\cal W}_R^{(k,l)}$ is the orthogonal complement
to ${\cal A}_R^{k-1}$ in ${\cal A}_R^l$ and
$${\cal A}_R^p=\bigoplus_{k=0}^p{\cal W}_R^{(k,k)}\eqno(75)$$
{\it For a fixed CC excitation level} $l$ {\it the subspace} ${\cal
W}_R^{(1,l)}$ {\it is 'the subspace of CC amplitudes'}.

Let us define the multiplication in ${\cal A}_R^p$ by putting
$$e_J^I\star e_{J'}^{I'}= \cases{(-1)^{|(J\cup I)\cap
{\Delta}_{(J'\cup I')}|}e_{J\cup J'}^{I\cup I'} &if $J\cap
J'=\emptyset$ and $I\cap I'=\emptyset$\cr 0 &if $J\cap J'\ne
\emptyset \ $ or $\ I\cap I'\ne \emptyset$\cr}.\eqno(76)$$ It is
easy to see that with such a multiplication the vector space
${\cal A}_R^p$ becomes a commutative and associative algebra with
$e_{\emptyset}^{\emptyset}$ as its identity (compare with
algebras studied by Paldus \cite{Paldus-1}).
Commutativity follows directly from Eq.(11), associativity from
the relation
$$|(J_1\cup I_1)\cap {\Delta}_{(J_2\cup I_2})|+|(J_1\cup J_2\cup
I_1\cup I_2)\cap {\Delta}_{(J_3\cup I_3)}|$$
$$\equiv |(J_2\cup I_2)\cap {\Delta}_{(J_3\cup I_3})|+|(J_1\cup I_1)\cap
{\Delta}_{(J_2\cup J_3\cup I_2\cup I_3)}| (mod\ 2).$$

{\it The structure introduced depends on the choice of the HF vacuum
vector and for a fixed MSO basis set there are exactly
${n\choose p}$ such (isomorphic) structures.}

As has already been mentioned, for vector space structure on
${\cal A}_R^p$ the change of  vacuum vector simply means change of
notation for the same basis determinants (see Eqs.(70) - (71)). In
contrast, the algebra structure defined by Eq.(76), is deeply
connected with the basic vacuum vector. For example,
$e_{\emptyset}^{\emptyset}(S)=|S\rangle $ is the idempotent of
algebra ${\cal A}_S^p$  whereas the same determinant $|S\rangle
=e_J^I(R)$ in algebra ${\cal A}_R^p$ is its nilpotent basis element.
Isomorphism between ${\cal A}_S^p$ and ${\cal A}_R^p$ is
induced, for example,  by a permutation  $\sigma$ of the MSO
index set $N$ such that $\sigma (S)=R$.

The relation
$${\cal W}_R^{(k_1,k_1)}\star {\cal
W}_R^{(k_2,k_2)}\subset {\cal W}_R^{(k_1+k_2,k_1+k_2)}\eqno(77)$$
where $k_1+k_2\le p$, shows that ${\cal A}_R^p$ is a graded
algebra. The subspace ${\cal W}_R^{(1,p)}$ is its maximal
nilpotent ideal and the algebra under discussion is just a direct
sum of the field of complex numbers and this nilpotent ideal
$${\cal A}_R^p=\mathbb C e_{\emptyset}^{\emptyset}\oplus {\cal
W}_R^{(1,p)}\eqno(78)$$ Note also that ${\cal A}_R^p$ is algebra
with involution induced by the complex conjugation.

It is well-known \cite {Exp} that in algebras of such type it is
easy to define {\it algebraically} both the exponential mapping
and its inverse. Namely,
$$exp:\tau\to \sum\limits_{k=0}^p\frac{{\tau}^k}{k!}\eqno(79)$$
is the exponential mapping ${\cal W}_R^{(1,p)}\to
e_{\emptyset}^{\emptyset}+{\cal W}_R^{(1,p)}$ and
$$log: e_{\emptyset}^{\emptyset}+\tau\to \sum\limits_{k=1}^p
(-1)^{k-1}\frac{{\tau}^k}{k}\eqno(80)$$ is the logarithmic
mapping $e_{\emptyset}^{\emptyset}+{\cal W}_R^{(1,p)}\to {\cal
W}_R^{(1,p)}$. Here
$${\tau}^k=\underbrace { \tau\star \ldots\star\tau}_k.\eqno(81)$$

For mappings (79) and (80) all classic relations hold true:
$$exp({\tau}_1)\star exp({\tau}_2)=exp({\tau}_1+{\tau}_2),\eqno(82a)$$
$$[exp(\tau)]^{-1}=exp(-\tau),\eqno(82b)$$
$$exp(log(e_{\emptyset}^{\emptyset}+\tau))=\tau,\eqno(82c)$$ and
$$log(exp(\tau))=\tau,\eqno(82d)$$ where $\tau, {\tau}_1, {\tau}_2$ are arbitrary elements from ${\cal
W}_R^{(1,p)}$.

If $\tau \in {\cal W}_R^{(1,l)}$ is a vector of CC amplitudes then
it is easy to show that the CI coefficients (56) are just the
components of the vector $exp(\tau)$ in algebra ${\cal A}_R^p$:
$$exp(\tau)=\sum\limits_{k=0}^p\sum\limits_{J\subset R\atop{I\subset N\backslash
R}}^{(k)}T_k^l(J,I)e_J^I,\eqno(83a)$$ where
$$T_k^l(J,I)=\langle exp(\tau)|e_J^I\rangle . \eqno(83b)$$
In fact, $exp(\tau)$ in the last equation for each fixed $k$   may
be replaced by the sum of the first $k$ terms in its expansion.
Eqs.(82) - (83) may be considered as a source of different relations
involving CC CI coefficients.

 {\it Thus, the Fock space equipped with the non-trivial structure of
commutative algebra, defined by Eq.(76), can be considered as a
natural domain for the CC and related approaches.}

Spectral theory in algebra ${\cal A}_R^p$ is very simple. Indeed,
let us write  arbitrary element $x \in {\cal A}_R^p$ as
$$x=x_{\emptyset}^{\emptyset}e_{\emptyset}^{\emptyset}+\tau$$
 where $\tau \in {\cal W}_R^{(1,p)}.$
Its spectrum is defined as the set of all $\lambda \in \mathbb C$
such that $(\lambda e_{\emptyset}^{\emptyset}-x)$ is not
invertible in ${\cal A}_R^p$. It is easy to see that $$(\lambda
e_{\emptyset}^{\emptyset}-x)^{-1}=\sum\limits_{k=0}^p\frac{{\tau}^k}{(\lambda
- x_{\emptyset}^{\emptyset})^{k+1}}={\cal R}(x,\lambda)\eqno(84)$$
exists for all $\lambda \ne x_{\emptyset}^{\emptyset}$. Thus,
spectrum of element $x\in {\cal A}_R^p$ is
$$Sp\ x=\{\langle x|e_{\emptyset}^{\emptyset}\rangle \}\eqno(85)$$
The mapping $\mathbb C\backslash
\{x_{\emptyset}^{\emptyset}\}\to {\cal A}_R^p $ defined by Eq.(84)
is called {\it the resolvent of element} $x\in {\cal A}_R^p$.

From Eq.(85) it follows that if element $x\in {\cal A}_R^p$
corresponds to normalized $p-$electron wave functions then its
spectrum belongs to the unit disk of the complex plane :
$$Sp\ x \subset {\bf D}=\{z\in \mathbb C:|z|\le 1\}$$

 In general case the structure of normed space
on ${\cal A}_R^p$ is not compatible with the algebra structure
defined by Eq.(76), that is the inequality
$$\|x\star y\|\le \|x\|\cdot \|y\|\eqno(86)$$
not necessarily holds true. For example, if $n=2$ and $p=1$, then
for $x=e_{\emptyset}^{\emptyset}+e_{\{1\}}^{\{2\}}$ the inequality
(86) is violated. Indeed, in this case $x\star
x=e_{\emptyset}^{\emptyset}+2e_{\{1\}}^{\{2\}}$ and $\|x\star
x\|=\sqrt{5}$ whereas $\|x\|^2=2$. Our hypothesis is \bigbreak
{\bf Hypothesis.} For sufficiently large $n$  inequality (86) is
satisfied for any $x,y\in  {\cal A}_R^p$. \bigbreak
 If this hypothesis is true then for values of $n$ of actual
 interest ${\cal A}_R^p$ is a commutative unital Banach algebra(see, e.g., \cite {Banach-1,
Banach-2}). In this case differential calculus of full value could
be developed on ${\cal A}_R^p$. It would be too restrictive to
consider only finite expansions of the type of Eqs.(79) - (80) and
it would be possible to define the exponential and logarithmic
mappings by their series and prove classic relations (82a) - (82b)
for {\it arbitrary} $x\in {\cal A}_R^p$. And even more, Banach
algebras are very closely related to algebras of analytic
functions of complex variables \cite {Banach-1, Banach-2} and this
also could be of interest for applications.

The basic idea of CC and related methods is to parametrize subsets
of ${\cal A}_R^p$ using elements('amplitudes') from ${\cal
W}_R^{(1,l)}$ where $l$ is the maximal excitation level.
General parametrization of such a type may be defined
as an injective mapping
$$\pi:{\cal W}_R^{(1,l)}\to {\cal A}_R^p\eqno(87)$$
such that element $\pi(\tau)$ is invertible for any $\tau \in
{\cal W}_R^{(1,l)}$. The last requirement guarantees that $\langle
\pi (\tau)|e_{\emptyset}^{\emptyset}\rangle \ne 0 $, or, in other
words, that the HF reference state appears with nonzero
coefficient in expansion of $\pi(\tau)$ for each $\tau$. It is
clear that the exponential mapping (79) is a parametrization (see
Eq.(82b)). It possesses the following characteristic property:
{\it for any excitation level} $l$ {\it the set} $exp({\cal
W}_R^{(1,l)})$ {\it is a multiplicative Abelian group}. This
property is referred to as 'size consistency' and is usually
considered as essential advantage of the exponential parametrization
over other ones.

Another parametrization used in the quadratic
configuration interaction method (QCI) \cite {Pople} is given by
$${\cal R}(\tau,1)=\sum\limits_{k=0}^p{\tau}^k=(
e_{\emptyset}^{\emptyset}-\tau)^{-1}\eqno(88a)$$ where
$$Q_k^l(J,I)=\langle {\cal R}(\tau,1)|e_J^I\rangle . \eqno(88b)$$
are the corresponding CI coefficients. The mapping inverse to
${\cal R}(\tau,1)$ is
$${\cal R}^{-1}:e_{\emptyset}^{\emptyset}+
\tau\to\sum\limits_{k=1}^p(-1)^{k-1}{\tau}^k\eqno(89)$$

The exponential and QCI parametizations are just special cases of
general polynomial parametrization
$$P_a:\tau\to \sum\limits_{k=0}^pa_k{\tau}^k,\eqno(90)$$
where $a=(a_0,a_1,\ldots,a_p)\in {\mathbb C}^{p+1}$. Element
$P_a(\tau)$ is invertible for any $\tau$ and the mapping (90) is
injective if and only if $a_0\cdot a_1\ne 0$. The inverse to $P_a$
may be written in the form
$$P_a^{-1}:a_0e_{\emptyset}^{\emptyset}+\tau \to
\sum\limits_{k=1}^p(-1)^{k-1}b_k{\tau}^k, \eqno(91)$$
where coefficients $b_k$ are easily obtained from the relation
$$\sum\limits_{k=1}^p(-1)^{k-1}b_k\left [P_a(\tau)-
a_0e_{\emptyset}^{\emptyset}\right ]^k=\tau.\eqno(92)$$
We have $b_1=\frac{1}{a_1}, b_2=\frac{a_2}{a_1^3}$, etc.

It is pertinent to mention that if the excitation level $l$ is
less than the number of electrons $p$ then any parametrization
covers only a part of the full CI space. If $x$ is some CI vector
and $\langle {\pi}^{-1}(x)|e_J^I\rangle =0$ for all $J,I$ with
$|J|=|I|>l$ then ${\pi}^{-1}(x)\in {\cal W}_R^{(1,l)}$, that is
this CI vector $x$ is covered by the parametrization used.

The main attractive feature of any polynomial parametrization
consists in possibility to get approximation to the lowest
eigenvector of Hamiltonian by iterative solution of non-linear system of
relatively small order
$$\langle e_J^I|H|P_a(\tau)\rangle = E \langle
e_J^I|P_a(\tau)\rangle ,\eqno(93) $$ where $1\le |J|=|I|\le l$.

In algebra ${\cal A}_R^p$ it is easy to derive relations
involving vector $\tau$ and energy value $E$ and different from the standard
CC type equations.
For example, let us suppose that QCI parametrization (88)
covers some eigenvector of the Hamiltonian $H$, that is there exists
$\tau \in {\cal W}_R^{(1,l)}$ such that
$$H{\cal R}(\tau,1)=E{\cal R}(\tau,1)\eqno(94)$$
It is easy to ascertain that the mapping
$${\cal R}_E^{-1}:E e_{\emptyset}^{\emptyset}+
\tau\to\sum\limits_{k=1}^p(-1)^{k-1}\frac{{\tau}^k}{E^k}\eqno(95)$$
is inverse to the mapping $\tau\to E {\cal R}(\tau,1)$.
Since, by assumption, Eq.(94) holds true, we can apply the transformation (95)
to the both sides of Eq.(94) to get
$$\sum\limits_{k=1}^p (-1)^{k-1}
\frac {\left [ H{\cal R}(\tau,1)-Ee_{\emptyset}^{\emptyset}\right
]^k}{E^k}= \tau .\eqno(96)$$ This relation may serve as a base for
Newton-Raphson type iterative algorithms for solution of Eq.(94).

 In algebra ${\cal A}_R^p$
the expression for the first order density operator for arbitrary
polynomial parametrization  takes the form:
$$p{\rho }_R^{CC}=
\sum\limits_{j\in R} \langle P_a(\tau)|\left
[\sum\limits_{k=0}^p\sum\limits_{J\subset R\backslash \{j\}\atop
{I\subset N\backslash R}}^{(k)}|e_J^I\rangle \langle e_J^I|\right
]|P_a(\tau)\rangle |j\rangle \langle j|$$
$$+\sum\limits_{j\in N\backslash R}\langle P_a(\tau)|\left [
\sum\limits_{k=1}^p\sum\limits_{J\subset
R}^{(k)}\sum\limits_{I\subset N\backslash R\backslash
\{j\}}^{(k-1)}|e_J^{I\cup \{j\}}\rangle \langle e_J^{I\cup
\{j\}}|\right ]|P_a(\tau) \rangle |j\rangle \langle j|$$
$$-\sum\limits_{i,j\in R\atop{(i\ne j)}}\langle P_a(\tau)|\left [\sum\limits_{k=1}\sum\limits_{J\subset
R\backslash \{i,j\}}^{(k-1)}\sum\limits_{I\subset N\backslash
R}^{(k)}(-1)^{{\gamma}_2} |e_{J\cup \{i\}}^I\rangle \langle
e_{J\cup \{j\}}^I|\right ]|P_a(\tau)\rangle |j\rangle \langle i|$$
$$-\sum\limits_{j\in R\atop{i\in N\backslash R}}\langle P_a(\tau)|\left [\sum\limits_{k=0}^{p-1}
\sum\limits_{J\subset R\backslash \{j\}\atop{I\subset N\backslash
R\backslash \{i\}}}^{(k)}(-1)^{{\gamma}_1+{\gamma}_2}
|e_J^I\rangle \langle e_{J\cup \{j\}}^{I\cup \{i\}}|\right ]
|P_a(\tau)\rangle |j\rangle \langle i|$$
$$+\sum\limits_{j\in N\backslash R\atop{i\in R}}\langle P_a(\tau)| \left [\sum\limits_{k=1}^p
\sum\limits_{J\subset R\backslash \{i\}\atop{I\subset N\backslash
R\backslash \{j\}}}^{(k-1)}(-1)^{{\gamma}_1+{\gamma}_2}|e_{J\cup
\{i\}}^{I\cup \{j\}}\rangle \langle e_J^I|\right
]|P_a(\tau)\rangle |j\rangle \langle i|$$
$$+\sum\limits_{i,j\in N\backslash R\atop{(i\ne j)}}\langle P_a(\tau)|\left [\sum\limits_{k=1}^p
\sum\limits_{J\subset R}^{(k)}\sum\limits_{I\subset N\backslash R
\backslash \{i,j\}}^{(k-1)} (-1)^{{\gamma}_2}|e_J^{I\cup
\{j\}}\rangle \langle e_J^{I\cup \{i\}}|\right ]|P_a(\tau)\rangle
|j\rangle \langle i|,\eqno(97)$$ where the expressions for
${\gamma}_1$ and ${\gamma}_2$ are given by Eq.(62a).

If $f$ and $g$ are differentiable mappings $\mathbb R \to {\cal
A}_R^p$ then it is easy to see that
$$\frac{d}{d\lambda}[f(\lambda)\star g(\lambda)]=
[\frac{d}{d\lambda}f(\lambda)]\star
g(\lambda)+f(\lambda)\star
[\frac{d}{d\lambda}g(\lambda)].\eqno(98)$$  In
particular,  for {\it real} vector $\tau$  of CC amplitudes
the following relations hold true
$$\frac{\partial}{\partial
t_J^I}{\tau}^k=k{\tau}^{k-1}\star e_J^I,\eqno(99a)$$
$$\frac{\partial}{\partial
t_J^I}exp(\tau)= exp(\tau) \star e_J^I,\eqno(99b)$$
$$\frac{d}{d\lambda}exp(\lambda \tau)=exp(\lambda \tau)\star \tau,\eqno(99c)
$$
and
$$\frac{d}{d\lambda}log(e_{\emptyset}^{\emptyset}+\lambda \tau)=
\frac{{\cal R}^{-1}(\lambda \tau ,1)}{\lambda}.\eqno(99d)$$

In concluding this section let us note that it is easy to define
algebra ${\cal
A}_{(R_{\alpha},R_{\beta})}^{(p_{\alpha},p_{\beta})}$
corresponding to orbital representation of $p-$electron sector of
the Fock space spanned by determinants with fixed spin projection
$M_S=\frac{1}{2}(p_{\alpha}-p_{\beta})$. The multiplication in
${\cal A}_{(R_{\alpha},R_{\beta})}^{(p_{\alpha},p_{\beta})}$ is
$$e_{(J_{\alpha},J_{\beta})}^{(I_{\alpha},I_{\beta})}\star
e_{(J'_{\alpha},J'_{\beta})}^{(I'_{\alpha},I'_{\beta})}= $$
$$= \cases{
(-1)^{\sum\limits_{\sigma}|(J_{\sigma}\cup I_{\sigma})\cap
{\Delta}_{(J'_{\sigma}\cup I'_{\sigma})}|}
e_{(J_{\alpha}\cup J'_{\alpha},J_{\beta}\cup J'_{\beta})}^
{(I_{\alpha}\cup I'_{\alpha},I_{\beta}\cup I'_{\beta})}
&if $J_{\sigma}\cap
J'_{\sigma}=\emptyset$ and $I_{\sigma}\cap I'_{\sigma}=\emptyset$\cr
0 &if $J_{\sigma}\cap J'_{\sigma}\ne
\emptyset \ $ or $\ I_{\sigma}\cap I'_{\sigma}\ne \emptyset$\cr}\eqno(100)$$
where $\sigma=\alpha,\beta$.

\bigbreak \bigbreak
\newpage
{\Large \bf \ Pointed Fock Spaces as Associative}

{\Large \bf \  Superalgebras}
\bigbreak \bigbreak

\bigbreak \bigbreak

 Let us describe possible extension of the multiplication rule (76)
to the full Fock space. To this end we consider {\it a pointed
Fock space} ${\cal A}_R=({\cal F}_N,|R\rangle)$ with the basis
$e_J^I(R)=(R\backslash J)\cup I\rangle$ where $J\subset R$ and
$I\subset N\backslash R$( on the whole $2^p\cdot 2^{n-p}=2^n$
basis vectors) and present ${\cal A}_R$ as a direct sum
$${\cal A}_R={\cal A}_R^{+} \oplus {\cal A}_R^{-}\eqno(101)$$
where subspaces ${\cal A}_R^{+}$ and  ${\cal A}_R^{-}$ are spanned
by basis vectors $e_J^I(R)$ with {\it even} and {\it odd} values
of $|J\cup I|$, respectively. Using the same definition (76) for
the product of {\it arbitrary} basis vectors, we come to
associative algebra with identity. From Eq.(11) if follows that
${\cal A}_R$ is {\it a skew-commutative (or super-commutative) }
algebra:
$$e_J^I(R)\star e_{J'}^{I'}(R)=(-1)^{|J\cup I|\cdot |J'\cup
I'|} e_{J'}^{I'}(R)\star e_J^I(R).\eqno(102)$$
We have
$${\cal A}_R^{+}\star {\cal A}_R^{+}\subset {\cal A}_R^{+},\
{\cal A}_R^{+}\star {\cal A}_R^{-}\subset {\cal A}_R^{-},\  {\cal
A}_R^{-}\star {\cal A}_R^{+}\subset {\cal A}_R^{-},\  {\cal
A}_R^{-}\star {\cal A}_R^{-}\subset {\cal A}_R^{+}, \eqno(103)$$
which means that ${\cal A}_R$ is ${\mathbb Z}_2-$graded
(super)algebra and that ${\cal A}_R^{+}$ is a subalgebra of ${\cal A}_R$.

Let us consider the pointed Fock space ${\cal A}_{\emptyset}$ that
is the Fock space where  {\it the 'absolute' vacuum state}
$|R\rangle =|\emptyset\rangle $  is selected. In this case we have
$$e_{\emptyset}^I(\emptyset)\star e_{\emptyset}^{I'}(\emptyset)=\cases{(-1)^{|I\cap
{\Delta}_{I'}|}e_{\emptyset}^{I\cup I'}(\emptyset) &if $I\cap
I'=\emptyset $\cr 0 &if $I\cap I'\ne \emptyset $\cr}\eqno(104)$$

It is interesting to compare the star product in ${\cal
A}_{\emptyset}$ with the standard exterior (wedge) product in the
Grassmann algebra ${\cal F}_N$. We have
$$|I\rangle \wedge |I'\rangle = \cases
{ (-1)^{\varepsilon(I,I')} |I\cup I'\rangle &if $I\cap
I'=\emptyset $ \cr 0 &if $I\cap I'\ne \emptyset $\cr }\eqno(105)$$
where $\varepsilon(I,I')$ is the number of pairs $(i,i')\in
I\times I'$ such that $i>i'$. It is easy to show (see Eq.(11))
that
$$\varepsilon(I,I')=|I'\cap{\Delta}_{I}|\equiv
|I\cap{\Delta}_{I'}|+|I|\cdot |I'| (mod \ 2)$$ and this means that
even in the case of algebra ${\cal A}_{\emptyset}$ the star
product is different from the standard exterior product in the
Grassmann algebra ${\cal F}_N$. For example, if $I=\{1,2,7\}$ and
$I'=\{4,5,6\}$ then ${\Delta}_{I'}=\{1,2,3,4,6\}$. We have
$$|I\cap {\Delta}_{I'}|=2$$
whereas
$$\varepsilon (I,I')=|\{(7,4),(7,5),(7,6)\}|=3.$$

Another possible extension of the multiplication rule (76) is
given by the relation
$$e_J^I\star e_{J'}^{I'}= \cases{(-1)^{|(J'\cup I')\cap
{\Delta}_{(J\cup I)}|}e_{J\cup J'}^{I\cup I'} &if $J\cap
J'=\emptyset$ and $I\cap I'=\emptyset$\cr 0 &if $J\cap J'\ne
\emptyset \ $ or $\ I\cap I'\ne \emptyset$\cr}.\eqno(106)$$

Both definition (76) and definition (106)  lead to the same
structure on the $p-$electron sector of the Fock space. Their
extension to the full Fock space gives, however, different
results. In particular, with the definition (106) algebra ${\cal
A}_{\emptyset}$ is identical to the Grassmann algebra.
 \bigbreak \bigbreak

{\Large \bf \  Computer Implementation of Star Product}

\bigbreak \bigbreak

 For two arbitrary vectors $x,y\in
{\cal A}_R^p$ we have
$$x\star y=x_{\emptyset}^{\emptyset}y+y_{\emptyset}^{\emptyset}x +
\sum\limits_{k=1}^p\sum\limits_{J\subset R \atop {I\subset
N\backslash R}}^{(k)} \left [
\sum\limits_{k_1=1}^{k-1}\sum\limits_{J_1\subset J\atop{I_1\subset
I }}^{(k_1)}(-1)^{k_1+|(J_1\cup I_1)\cap {\Delta}_{(J\cup
I)}|}x_{J_1}^{I_1}y_{J\backslash J_1}^{I\backslash I_1}\right
]e_J^I .\eqno(107)$$

Methods of calculation of this rather complicated product are
based on two fundamental combinatorial  notions: listing and
ranking. Listing means generation of combinatorial objects
(subsets of a given index set in our case) in some fixed ordering.
Ranking algorithms allow to set up arrays indexed by subsets
(vectors of algebra ${\cal A}_R^p$ in our case). In other words,
rank of a subset is the position this subset occupies in a given
ordering.

There exists simple formula for calculation of subset rank under
lexical ordering of the set of all $k-$element subsets of the
index set $N$ (see, e.g., \cite {RND}):
$$lex_N(J)={n\choose k}-\sum\limits_{j=1}^n
{\zeta}_{j,J}{{n-j}\choose {k-|J\cap {\Delta}_{\{j\}}|}+1}.
\eqno(108)$$

In algebra ${\cal A}_R^p$ components of arbitrary vector $x$ are
indexed by pairs of subsets $(J,I)$ with $J\subset R$ and
$I\subset N\backslash R$. Let us put
$$lex_R(J)=lex_{r^{-1}(R)}(r^{-1}(J))\eqno(109a)$$
and
$$lex_{N\backslash R}(I)=lex_{s^{-1}(N\backslash R)}(s^{-1}(I))\eqno(109b)$$
where $R=r_1<\ldots<r_p$ and $N\backslash R=s_1<\ldots<s_{n-p}$.
Vector $x\in {\cal A}_R^p$ has its components indexed as
$$(J,I)\to (lex_R(J)-1){{n-p}\choose k}+lex_{N\backslash R}(I).\eqno(110)$$
In our opinion ranking and listing algorithms required for
evaluation of products $x\star y$ should not necessarily be
closely related. For example, we can use Gray codes generation
algorithms \cite {NW,Kreher} and lexical indices (110).

\bigbreak

{\Large \bf \ Conclusion}

\bigbreak \bigbreak

Advanced technique of manipulations with phase prefactors arising
in many problems of quantum chemistry is demonstrated. On the base
of this technique three methods of evaluation of CC CI
coefficients and CC density matrices are suggested.

The first method (see Eq.(56)) requires rather complicated
combinatorial algorithms for set partitions generation and
permutations of multisets (see Appendix). The success of its
implementation strongly depends on the program code efficiency.

The second method (see Eq.(68)) is based on the recurrence
relations for CC CI coefficients and its efficiency is closely
related both to the efficiency of recursive algorithms in
programming language used and available hardware.

The third method employs new algebraic structure revealed on the
Fock space. The efficiency of implementation of a rather special
product of two vectors determines the overall efficiency of any
algorithms using this multiplication rule.

The first two methods do not require full size CC CI vector to be
kept in fast memory but they may be rather time consuming. The
third method can be programmed in a very efficient way but at
least two CC CI vectors should be stored.

\bigbreak \bigbreak
\newpage
{\Large \bf \ Appendix: RG strings and set partitions}

\bigbreak \bigbreak

All information contained in this Appendix is well-known and is
given here only to show how direct evaluation of expressions of
the type of Eq.(56) can be performed in non-recursive way.
Standard references are \cite {NW,Kreher}. A lot of interesting
information may also be found on Frank Ruskey site \cite {Ruskey}.

 Let $\{J_i\}_{i=1}^{\mu}$ be a set partition of the set
$N=\{1,2,\ldots,n\}$ having its blocks listed in accordance to the
following rule: block $J_i$ contains the minimal element of the
set $N\backslash ({\bigcup }_{j=1}^{i-1}J_j )$. In particular,
$J_1\ni 1$. In this Appendix the term 'set partition' means 'block
ordered set partition' unless otherwise stated.

{\bf Definition.} Restricted Growth (RG) string of length $n$ is
an integral vector $a=(a_1,a_2,\ldots,a_n)$ such that
$$a_1\le a_i\le \max_{1\le j<i}\{a_j\}+1\eqno(1)$$
where $a_1=1$.

With each RG string it is convenient to associate the following
vector
$$\nu=({\nu}_1,\ldots,{\nu}_n)\eqno(2)$$
where ${\nu}_i=\max\{a_1,\ldots,a_i\}$.

 {\bf Lemma.} There exists a bijection between the set of all
RG strings of length $n$ and the set of all set partitions of $N$.

{\bf Proof.} See \cite {NW,Kreher}.

Non-recursive generation of all RG strings in {\it lexical} order
can be described as follows. We start with the minimal RG string
$$a_{min}=(1,1,\ldots,1)\eqno(3)$$
Let us suppose that on some step we have current RG string
$$(1,\ldots,a_m,a_{m+1},\ldots,a_n)\eqno(4)$$
where $|J_{a_m}|>1$ and $|J_{a_{m+1}}|=\ldots =|J_{a_n}|=1$. It is
easy to see that in RG string (4) $a_{m+1}>{\nu}_m$. Indeed, in
the opposite case the smallest element of $J_{a_{m+1}}$ would be
not greater than $m$ but this is in contradiction with the
assumption that $J_{a_{m+1}}$ is a block of size 1 containing
index $m+1$. As a result, the current RG string (4) takes
actually the form
$$(1,\ldots,a_m,{\nu}_m+1,\ldots,{\nu}_m+n-m)\eqno(5)$$
The corresponding set partition with ${\nu}_m+n-m$ blocks is
$$\{J_1,\ldots,J_{a_m},\ldots,J_{{\nu}_m},\{m+1\},\ldots,\{n\}\}$$

The next (in lexical order) RG string is
$$(1,\ldots,a_m+1,1,\ldots,1)\eqno(6)$$
In going from RG string (5) to RG string (6) the size of the
block $J_1$ increases by $n-m$ and the current number of blocks
decreases by the same value and becomes equal to ${\nu}_m$ if
$a_m\ne {\nu}_m$. If $a_m = {\nu}_m$ then RG string (6)
corresponds to the set partition with ${\nu}_m+1$ blocks.

Generation of RG strings should be terminated when the number of
current blocks becomes equal to $n$ that corresponds to the
maximal RG string
$$a_{max}=(1,2,\ldots,n)\eqno(7)$$
and the set partition with exactly $n$ blocks $J_i=\{i\}$.

 We are interested  in generation of set
partitions with the additional restriction on block sizes. Namely,
for any partition of $N$ the size of each its block should not be
greater than the maximal excitation order $l$. This implies, in
particular, that with such a restriction the number of blocks can
not be less than
$$k=\cases{[\frac {n}{l}] &if $n\equiv 0 (mod \ l)$\cr
           [\frac {n}{l}]+1 &if $n\not\equiv 0 (mod \ l)$\cr}\eqno(8)$$

With {\it lexical} ordering of admissible RG strings the minimal
one is
$$a_{min}=\cases {(\underbrace{1,\ldots,1}_l,\underbrace{2,\ldots,2}_l,\ldots,\underbrace
{k,\ldots,k}_{l}) &if $n\equiv 0 (mod \ l)$\cr
(\underbrace{1,\ldots,1}_l,\underbrace{2,\ldots,2}_l,\ldots,
\underbrace{k,\ldots,k}_{n-l\cdot[\frac {n}{l}]})&if $n\not\equiv
0 (mod \ l)$\cr}\eqno(9)$$ and the maximal one is given by
Eq.(7).

Let us suppose that the current RG string is of the form of
Eq.(5) and that the corresponding vector of block sizes is
$$s=(|J_1|,\ldots,|J_{a_m}|,\ldots,|J_{{\nu}_m}|,1,\ldots,1).\eqno(10)$$
 Since each block size should not exceed $l$,
it is necessary to scan vector (10) from the position $a_{m}+1$
till ${\nu}_m$ to find the smallest index $\bar {a}_m$ such that
$|J_{\bar{a}_m}|<l$. If such an index is not found then we put
$\bar { a}_m={\nu}_m+1$. The next string is
$$(1,\ldots,\bar {a}_m,\underbrace{1,1,\ldots ,}_{{\kappa}_1}\underbrace{2,2,\ldots,}_{{\kappa}_2}\ldots )$$
where
${\kappa}_1=min\{l-|J_1|,n-m\},{\kappa}_2=min\{l-|J_2|,n-m-{\kappa}_1\}$,
etc. If after sorting blocks
$J_1,\ldots,J_{{\nu}_m}(J_{{\nu}_m+1})$ there are still $\kappa
=n-m-\sum\limits_{i=1}{\kappa}_i>0$ unfilled positions in the
string under consideration, then we should fill them by
${\delta}_1=min\{l,\kappa\}$ integers ${\nu}_m+1 ({\nu}_m+2)$,
etc.

 The next algorithm required is the algorithm of
generation of all set partitions (not necessarily block ordered
ones) corresponding to a given vector $s=(s_1,\ldots,s_{\mu})$ of
block sizes. But this problem is equivalent to the problem of
generation of multiset permutations and the corresponding
non-recursive lexical algorithm is easily obtained by modification
of, say, algorithm 5.1 from the book \cite {RND}.

\bigbreak \bigbreak

{\Large \bf \ Acknowledgments} \bigbreak \bigbreak

The author  gratefully acknowledges the Russian Foundation for
Basic Research (Grant 03-03-32335a) for financial support of the
present work. Special thanks to Prof. A. V. Titov for bringing to
author's notice the problem of CC 1-density operator construction
as well as for useful and inspiring discussions.
 \bigbreak

\bigbreak

\end{document}